\documentclass[fleqn,usenatbib]{mnras}
\usepackage{newtxtext,newtxmath}
\usepackage[T1]{fontenc}
\usepackage{float}
\DeclareRobustCommand{\VAN}[3]{#2}
\let\VANthebibliography\thebibliography
\def\thebibliography{\DeclareRobustCommand{\VAN}[3]{##3}\VANthebibliography}
\usepackage{graphicx}	
\usepackage{amsmath}	

\title[]{The impact of ultraviolet heating and cooling on the dynamics and observability of lava planet atmospheres}
\author[Nguyen et al.]{
T. Giang Nguyen,$^{1}$\thanks{E-mail: giang@yorku.ca}
Nicolas B. Cowan,$^{2}$
Raymond T. Pierrehumbert,$^{3}$
Roxana E. Lupu,$^{4}$
and John E. Moores$^{1}$
\\
$^{1}$Centre for Research in Earth and Space Sciences, York University, 4700 Keele St, Toronto, ON M3J 1P3, Canada\\
$^{2}$Department of Earth and Planetary Sciences, and Department of Physics, McGill University, 3550 Rue University, Montr\'eal, QC H3A 2A7, Canada\\
$^{3}$Department of Physics, University of Oxford, Oxford, OX1 3PU, United Kingdom\\
$^{4}$BAER Institute/NASA Ames Research Center, Moffet Field, CA 94035, USA
}

\date{Accepted XXX. Received YYY; in original form ZZZ}

\pubyear{2021}
\graphicspath{{figures/}}
\begin{document}
\label{firstpage}
\pagerange{\pageref{firstpage}--\pageref{lastpage}}
\maketitle

\begin{abstract}
Lava planets have non-global, condensible atmospheres similar to icy bodies within the solar system. Because they depend on interior dynamics, studying the atmospheres of lava planets can lead to understanding unique geological processes driven by their extreme environment. Models of lava planet atmospheres have thus far focused on either radiative transfer or hydrodynamics. In this study, we couple the two processes by introducing ultraviolet and infrared radiation to a turbulent boundary layer model. We also test the effect of different vertical temperature profiles on atmospheric dynamics. Results from the model show that UV radiation affects the atmosphere much more than IR. UV heating and cooling work together to produce a horizontally isothermal atmosphere away from the sub-stellar point regardless of the vertical temperature profile. We also find that stronger temperature inversions induce stronger winds and hence cool the atmosphere. Our simulated transmission spectra of the bound atmosphere show a strong SiO feature in the UV that would be challenging to observe in the planet's transit spectrum due to the precision required. Our simulated emission spectra are more promising, with significant SiO spectral features at 4.5 and 9 $\mu$m that can be observed with the James Webb Space Telescope. Different vertical temperature profiles produce discernible dayside emission spectra, but not in the way one would expect.
\end{abstract}




\section{Introduction}
\subsection{Background}
Lava planets are rocky exoplanets that orbit close to their star. Being tidally-locked into synchronous rotation, these planets have large surface temperature contrasts between the permanent dayside and nightside. On the dayside, the stellar flux is intense enough to vaporize rocks, generating a thin atmosphere of mineral vapour \citep{schaefer2009chemistry}. On the nightside, the temperature drops and the atmosphere may collapse  back onto the surface as has been predicted for volatiles on M-Earths \citep{wordsworth2015atmospheric}. In order to qualify as a lava planet, a world must have a density consistent with rocky bulk composition, and the surface temperature must exceed the solidus for silicates; examples include CoRoT-7b \citep{leger2011extreme}, Kepler-10b \citep{batalha2011kepler}, and K2-141b \citep{malavolta2018ultra}.

Dynamics of the partial atmospheres on lava planets are similar to that of Io \citep{tsang2016collapse} or Pluto \citep{gladstone2019new} where pressure can vary significantly in response to huge changes in temperature. As such, icy bodies in the solar system are counter-intuitively analogous to lava planets and studying one can help to understand the other. The atmosphere on a lava planet is in vapour equilibrium with the magma ocean, so atmospheric and interior dynamics are intertwined \citep{kite2016atmosphere}. Therefore, atmospheric detection and characterization of a lava planet may reveal the bulk composition of its rocky mantle. 

\cite{ito2015theoretical} used radiative transfer simulation to show that a silicon monoxide atmosphere near the substellar point can be much hotter than the surface and exhibit a strong inversion (negative lapse rate). Infrared spectroscopy should reveal dayside SiO emission features. However, such 1D radiative transfer simulations neglect atmospheric circulation.

In contrast, \cite{nguyen2020modelling} focused on the hydrodynamical flow of the silicate atmosphere, with an emphasis on the horizontal advection of heat and the spatial extent of the atmosphere. We had modelled an optically thin silicate atmosphere that did not absorb or emit any radiation. This made it difficult to simulate observables such as emission and transmission spectra. By construction, the optically thin atmosphere does not absorb stellar radiation and thus was everywhere colder than the surface. A temperature inversion as envisioned by \cite{ito2015theoretical} might affect the atmospheric dynamics. Therefore, the next step is to combine radiative transfer and atmospheric circulation to better understand lava planets atmospheric dynamics and more faithfully simulate observations.

\subsection{Objectives}

\cite{castan2011atmospheres} and \cite{nguyen2020modelling} use the same formulation --- they are insightful for atmospheric dynamics but both assume a transparent atmosphere. Since SiO vapour absorbs infrared and ultraviolet radiation very well, we aim to eliminate this approximation by adding radiative transfer to the evaporation-driven hydrodynamics. Hydrodynamical simulations have also assumed a dry adiabatic vertical temperature profile, contrary to the 1D radiative transfer simulations of \cite{ito2015theoretical}. We therefore also test a wider range of vertical temperature profiles to harmonize with the radiative transfer simulations.

The resulting model couples radiative transfer to atmospheric circulation, something yet to be done for lava planets. As an ancillary benefit, they also allow us to more faithfully predict observables. By calculating the transmission and emission spectra, we simulate JWST observations. We continue using K2-141b as our poster child as it currently has the highest signal-to-noise ratio among known lava planets making it the best target for atmospheric detection \citep{zieba2022k2} and is one of the first planets that JWST will look at \citep{dang2021hell}.


The next section, Section \ref{sec_Theory}, describes the model's formulation: introduction of the general turbulent boundary layer (\ref{subsec_TBL}), simplification of the general equations to 1D (\ref{subsec_hydro}), and the implementation of radiative transfer (\ref{subsec_RT}). Section \ref{sec_results} presents results from the model. Section \ref{sec_disc} discusses the impact of UV radiation (\ref{subsec_UV}) and different vertical temperature profiles (\ref{subsec_TP}) on atmospheric dynamics, as well as presenting simulated transit spectra (\ref{subsec_transpec}) and eclipse spectra (\ref{subsec_emispec}). We conclude in Section \ref{sec_conc}.

\section{Theory}
\label{sec_Theory}

\subsection{The turbulent boundary layer}
\label{subsec_TBL}

Following the formulation of \cite{ingersoll1985supersonic}, the atmosphere consists of a turbulent boundary layer that behaves similarly to the shallow-water equations. The surface pressure is assumed to be in fast equilibration with the magma ocean. As the flow is assumed to be in steady-state, we have the following general equations describing the conservation of mass, momentum, and energy respectively:
\begin{equation}
    \nabla \cdot (\rho h V) = m \ E,
    \label{eq_massg}
\end{equation}
\begin{equation}
    \nabla \cdot (\rho h V^2) = -\nabla \int_z P dz + \tau,
    \label{eq_momg}
\end{equation}
\begin{equation}
    \nabla \cdot\bigg(\rho h V \left(\frac{V^2}{2} + C_p T\right)\bigg) = Q,
    \label{eq_eneg}
\end{equation}
where $\rho$ is the air density, $h$ is the atmospheric column thickness, $V$ is wind velocity, $m$ is the mass per molecule, $P$ is the pressure, $\tau$ is the surface drag, $C_p$ is the heat capacity, $T$ is the atmospheric temperature, and $Q$ is the net energy flux for the boundary layer. The evaporation/outgassing rate $E$, is calculated via:
\begin{equation}
    E = \frac{P - P_v(T_s)}{m \sqrt{2 \pi R T_s}},
\end{equation}
\noindent where $R$ is the gas constant, $T_s$ is the surface temperature, and $P_v$ is the temperature-dependent saturation vapour pressure and can be found in \cite{miguel2011compositions}.

The surface drag, $\tau$, is calculated by:
\begin{equation}
    \tau = -\rho V w,
\end{equation}
\noindent where $w$ is the transfer coefficient defined by \cite{ingersoll1985supersonic} and is further parameterized by the mean flow velocity, $V_e = m E/\rho$, and the eddy velocity, $V_d = V_*^2/V$. The frictional velocity, $V_*$, is implicitly solved from the experimental formulation of turbulent flow over smooth flat plates:
\begin{equation}
    V = 2.5 V_* \log\bigg(\frac{9.0 \ V_* \ H \rho}{2\eta}\bigg),
\end{equation}
\noindent where $H$ is the atmospheric scale height and $\eta$ is the dynamic viscosity. We can now compute the transfer coefficient $w$ for the two cases where there is net evaporation ($V_e > 0$) and net deposition ($V_e \leq 0$):
\begin{subequations}
\begin{align}
   w &= \frac{2V_d^2}{V_e+2V_d} && (V_e > 0) \\
   w &= \frac{V_e^2-2 V_d V_e + 2 V_d^2}{-V_e + 2V_d} && (V_e \leq 0).
\end{align}
\end{subequations}

\begin{figure}
    \centering
    \includegraphics[width=\columnwidth]{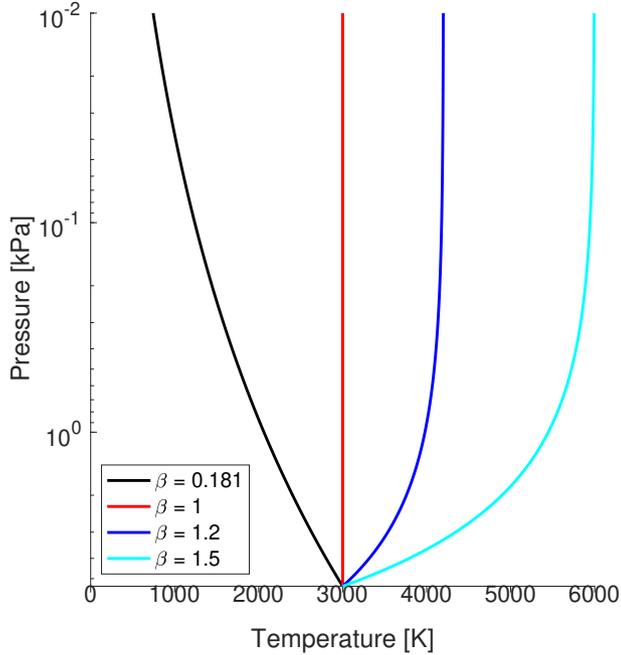}
    \caption{T-P profiles at the substellar point adopted by \protect\cite{nguyen2020modelling} for different $\beta$. The black line represents the adiabatic profile. The red line represents isothermal and the two blue lines represent profiles with a temperature inversion.}
    \label{fig_PTprofile}
\end{figure}

As the flow is assumed to be turbulent, $V$ can be vertically uniform. To solve the momentum equation, $P$ is vertically integrated with height by using the pressure and temperature at the top of the boundary layer, $P_*$ and $T_*$ respectively:

\begin{equation}
    \int_0^{\infty} P dz = \int_0^{P_*} \frac{\Phi}{g} dP = \frac{C_p}{g} \int_0^{P_*} (T_* - T) dP,
    \label{eq_Pint}
\end{equation}

\noindent where $\Phi$ is the geopotential. The change in $\Phi$ is the work done as the vaporized particles ascend from the surface to the top of the boundary layer; this can be expressed as $C_p(T_*-T)$. In order to close the problem, we must assume a vertical temperature profile. For an adiabatic profile, we have $T = T_*(P/P_*)^{R/C_p}$. Substituting this into the integral in Eq. \ref{eq_Pint} yields:

\begin{equation}
    \frac{C_p}{g} \bigg(T_* P_* - T_* P_* \bigg(\frac{1}{R/C_p + 1}\bigg)\bigg) = \frac{C_p T_* P_* \beta}{g},
\end{equation}

\noindent where $\beta = R/(R+C_p)$ for an adiabatic profile. For an isothermal atmosphere, on the other hand, $\beta = 1$ . Likewise an atmosphere with a temperature inversion would have $\beta$ greater than 1 (Figure \ref{fig_PTprofile} shows temperature -- pressure profiles for different $\beta$). The isothermal and inverted profiles are more in line with the temperature profile predicted by \cite{ito2015theoretical}. We will test these three different profiles and see their effects on the dynamics and observability of the K2-141b's atmosphere.

Because of our defined shallow turbulent layer, the vertical integration of pressure, temperature, and wind speed allow us to evaluate these state variables at the layer's boundary. Vertical integration requires an assumption of the temperature profile which we parameterized with $\beta$. The calculations become akin to the shallow-water equations where height is fixed within the column.

\subsection{Hydrodynamics}
\label{subsec_hydro}

We must now reduce the general equations \ref{eq_massg}-\ref{eq_eneg} to a more manageable set of calculations. As K2-141b is tidally locked, we can impose axial symmetry and the problem reduces to 2 dimensions: height and angular distance from the substellar point $\theta$. Assuming a shallow turbulent boundary layer further reduces height dependence. The problem becomes 1D where we solve for pressure, temperature, and wind velocity with respect to $\theta$ at the top of the turbulent layer. The general equations describing the conservation of mass, momentum, and energy become:
\begin{equation}
	\frac{1}{r \ \sin(\theta)}\frac{d}{d\theta}\bigg(\frac{V_* \ P_* \ \sin(\theta)}{g}\bigg) = m \ E,
	\label{eq_mass}
\end{equation}
\begin{equation}
	\frac{1}{r \ \sin(\theta)}\frac{d}{d\theta}\bigg(\frac{(V_*^2 + \beta \ C_p \ T_*)P_* \ \sin(\theta)}{g}\bigg) = \frac{\beta \ C_p \ T_* \ P_*}{g \ r \ \tan(\theta)} + \tau,
	\label{eq_mom}
\end{equation}
\begin{equation}
	\frac{1}{r \ \sin(\theta)}\frac{d}{d\theta}\bigg(\frac{(V_*^2/2 + \ C_p \ T_*)V_* \ P_* \ \sin(\theta)}{g}\bigg) = Q,
	\label{eq_en}
\end{equation}
where K2-141b's radius, $r$, and surface gravity, $g$, are $9.62\times10^6$ m and $21.8$ m  s$^{-2}$ respectively.

A limitation of the model is that the atmosphere's composition must be pure. Due to its volatility, sodium has been the prime outgassing candidate for previous lava planet modelling papers \citep{castan2011atmospheres,kite2016atmosphere}. However, Na is relatively scarce in the crust, and risks being lost to space \citep{ito2021hydrodynamic} or cold trapped on the nightside \citep{nguyen2020modelling}, so we argue instead for a silicate atmosphere due to the abundance of SiO$_2$ in the crust. Both SiO$_2$ and SiO can be evaporated into the atmosphere but the saturated vapour pressure of SiO$_2$ is several orders of magnitude lower \citep{schaefer2009chemistry}. Therefore, the atmosphere is likely dominated by SiO and its relevant parameters are: $m = 7.32\times10^{-26}$~kg~molecule$^{-1}$, $C_p = 851$~J~K$^{-1}$~kg$^{-1}$, $R = 188$~J~K$^{-1}$~kg$^{-1}$.

The system of ordinary differential equations, Eq.~\ref{eq_mass}-\ref{eq_en}, can be solved numerically as a boundary value problem. To find a solution, we use the shooting method where we test different sublimation rates at the substellar point to see which value leads to steady-state flow. When including radiative transfer, as described below, we must also test initial energy flux because evaporation is no longer the sole driving force of the flow.

\subsection{Radiative transfer}
\label{subsec_RT}

\begin{figure}
	\centering	
	\includegraphics[width=\columnwidth]{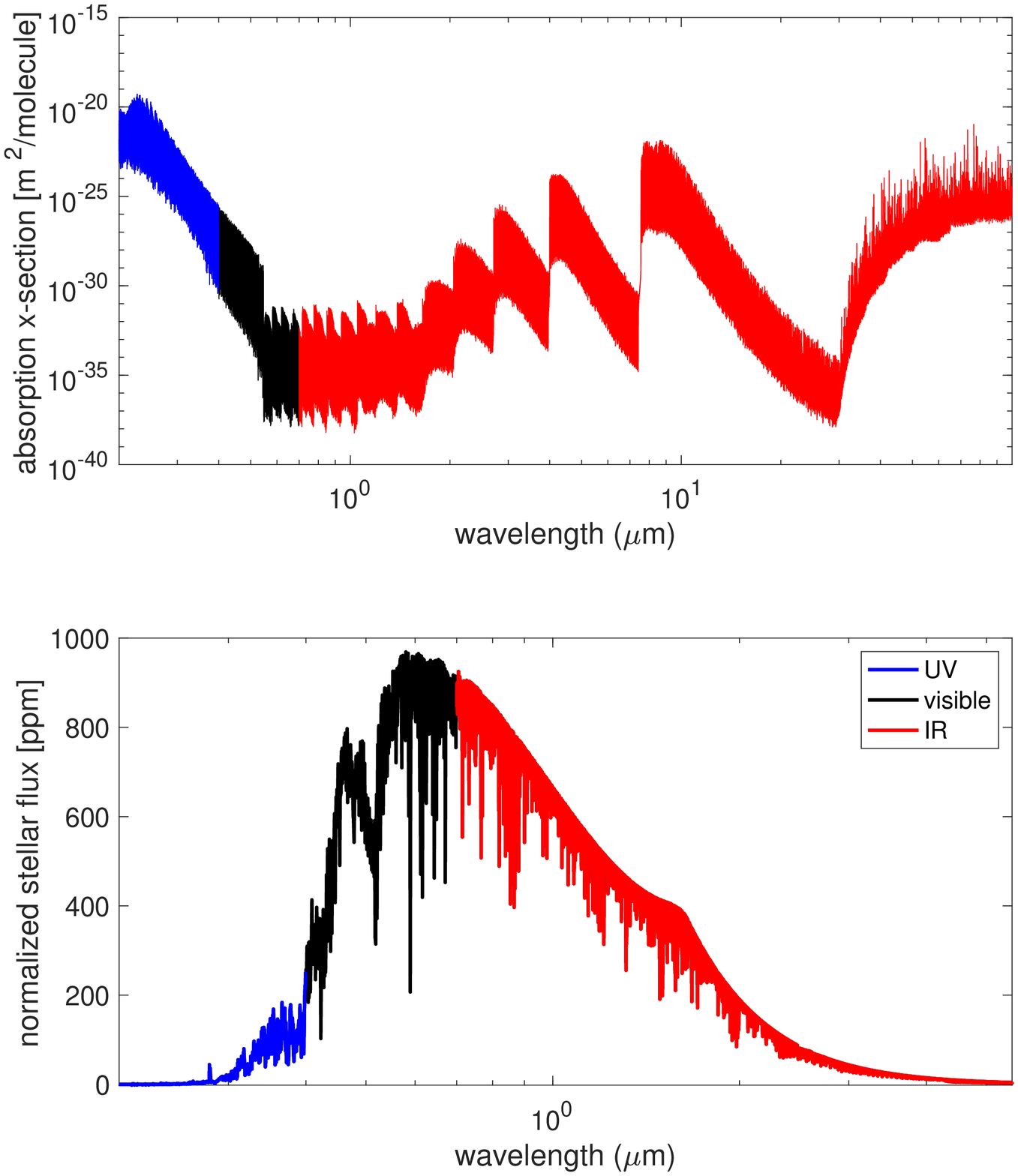}
	\caption{\emph{Top}: absorption cross-section of SiO at 1~kPa and 2800~K. The UV data are from \protect\cite{kurucz1992atomic} while the remainder are from \protect\cite{barton2013exomol}. Broadening parameters are approximated based on HITRAN data for similar diatomic molecules, assuming an H$_2$/He background atmosphere \protect\citep{gordon2017hitran2016}; although this assumption is hard to justify, this is currently the best source of SiO absorption cross-section at high temperatures. \emph{Bottom}: stellar spectrum of HD~5512 \citep{france2016muscles}, a close analogue of K2-141.}
	\label{fig_SiOspectrum}
\end{figure}

To implement radiative transfer in our model, we need the wavelength-dependent absorptivity of SiO and the stellar spectrum, both shown in Fig. \ref{fig_SiOspectrum}.  SiO absorbs strongly in the ultraviolet and relatively weakly in the infrared. Since SiO has negligible optical absorption, we only consider the IR and UV bands for heating and cooling of the atmosphere. 

Implementing radiative transfer starts with determining the absorptivity, or equivalently the emissivity, $\epsilon$, in each waveband:
\begin{equation}
    \epsilon = 1 - e^{-\tau_d},
\end{equation}
where $\tau_d$ is the optical depth. This depth is determined using the surface pressure:
\begin{equation}
    \tau_d = xP_*/mg,
\end{equation}
where $x$ is the absorption cross-section of SiO in the IR or UV: $x_{\rm IR} = 10^{-29} $ ~m$^2$/mole and $x_{\rm UV} = 10^{-22} $ ~m$^2$/mole.

With emissivity and absorptivity accounted for, we can now define the atmospheric heating $Q$ as:
\begin{equation}
	Q = Q_{\rm sens} + F_{\rm IR} + F_{\rm UV} + F_{\rm surf} - 2 F_{\rm RC},
	\label{eq_newQ}
\end{equation}
where $Q_{\rm sens}$ is the sensible heating as described by \cite{ingersoll1985supersonic}, $F_{\rm IR}$ is the atmospheric absorption of stellar IR, $F_{\rm UV}$ is the atmospheric absorption of stellar UV, $F_{\rm surf}$ is the blackbody radiation emitted by the surface and subsequently absorbed by the atmosphere, and $F_{\rm RC}$ is the radiative cooling of the atmosphere. The radiative heating terms for the atmosphere are:
\begin{equation}
F_{\rm IR} = \epsilon_{\rm IR}  \ C_{\rm IR} \  F_*,
\label{1}
\end{equation}
\begin{equation}
	F_{\rm UV} = \epsilon_{\rm UV}  \ C_{\rm UV} \  F_*,
	\label{2}
\end{equation}
\begin{equation}
	F_{\rm surf} = \epsilon_{\rm IR} \sigma T_s^4.
	\label{3}
\end{equation}

In the equations above, $F_*$ is the incident stellar flux, which is a function of $\theta$ calculated following \cite{kopal1954photometric} to account for the non-negligible angular size of the star as seen from the planet. For the stellar spectrum, we use the star HD~5512 as it is a close analogue of K2-141; $C_{\rm IR} = 0.532$ and $C_{\rm UV} = 0.009$ are the fraction of stellar flux in the IR and UV range, respectively \citep{france2016muscles}. We ignore stellar UV absorption via photo-dissociation of SiO because, as we show below, absorptivity and emissivity in the UV is $\sim$1 for most the dayside. Since nearly all of the stellar UV has been absorbed, any processes that can further absorb UV are negligible. Furthermore, significant photo-dissociation of SiO occurs at wavelengths less than 0.14 $\mu$m \citep{jolicard1997photodissociation} and stellar flux in this range is negligible.

For radiative cooling, only considering IR emission would yield $F_{\rm RC} = \epsilon_{\rm IR} \sigma T_*^4$. However, since the absorption cross-section of the atmosphere is several orders of magnitude larger in the UV than the IR, UV emission becomes non-negligible. Therefore we split the $\epsilon  \sigma T_*^4$ term using definite integrals of the Planck function, $B$, over IR and UV wavelengths, $\lambda$, and the equation for $F_{\rm RC}$ becomes:
\begin{equation}
	F_{\rm RC}(T) = \pi \bigg(\epsilon_{\rm IR} \int_{\lambda_{\rm IR}} B(\lambda,T) d\lambda + \epsilon_{\rm UV} \int_{\lambda_{\rm UV}} B(\lambda,T) d\lambda\bigg).
	\label{5}
\end{equation}

A computationally efficient way to evaluate the Planck integral can be found in the Appendix. The surface albedo is assumed to be 0 and this is justified as quenched glass and liquid lava surfaces are expected to have very low albedo \citep{essack2020low}. We account for the greenhouse effect by radiatively coupling the atmospheric and surface temperatures:
\begin{equation}
	\sigma T_s^4 = F_* - F_{\rm IR} - F_{\rm UV} + F_{\rm RC}(T).
\end{equation}

\section{Results}
\label{sec_results}

\begin{figure}
	\centering	
	\includegraphics[width=\columnwidth,height=\textheight,keepaspectratio]{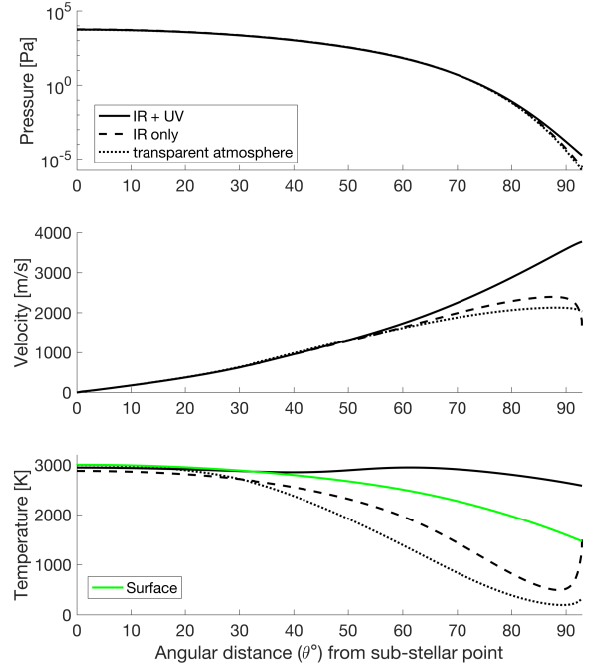}
	\caption{Atmospheric pressure, wind velocity and temperature from 1D hydrodynamic + radiative transfer simulations of K2-141b assuming an adiabatic vertical temperature-pressure profile. In each panel the dotted line represents the transparent atmosphere \citep[no radiative transfer, as in][]{nguyen2020modelling}, the dashed line represents the simulation with only IR radiative transfer, and the solid line is the simulation incorporating both IR and UV radiative transfer. The IR-only radiative scheme is similar to the transparent case but the scheme with UV radiation produces stronger winds and a much hotter atmosphere. It can be seen that the temperature for the IR-only case rises significantly at high $\theta$. This is caused by viscosity which slow winds and raises the temperature, but is only effective at low temperatures. However, the transparent atmosphere is just as cold but does not heat up. This is because it has significantly less pressure than its IR-only counterpart and there is not enough atmospheric mass for friction to take effect.}
	\label{fig_PVTold}
\end{figure}

Starting with an adiabatic profile ($\beta = 0.18$), we first only include IR absorption and emission. The results did not differ much from a transparent atmosphere aside from having slightly stronger winds and warmer atmospheric temperatures. Atmospheric temperatures are still always cooler than the surface (cf.\ dotted and dashed lines in Fig.~\ref{fig_PVTold}). Pressure almost never changes regardless of the radiative schemes or temperature profiles used as it cannot deviate far from the local saturation vapour pressure.

\begin{figure}
	\centering	
	\includegraphics[width=\columnwidth,height=\textheight,keepaspectratio]{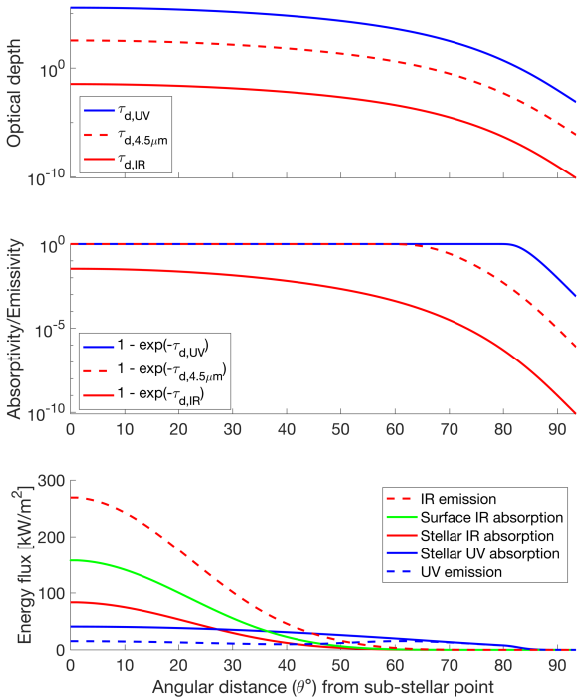}
	\caption{Radiative properties of the adiabatic atmosphere. \emph{Top}: vertical optical depth of the SiO atmosphere. \emph{Middle}: vertically integrated absorptivity (equivalently emissivity). \emph{Bottom}: radiative fluxes absorbed and emitted by the atmosphere. The emissivity in the UV is $\sim$1 for most of the dayside while emissivity in the IR drops exponentially away from the substellar point. While UV heating and cooling are relative small near the sub-stellar point, they dominate the radiation budget at $\theta>50^\circ$.}
	\label{fig_epsilon}
\end{figure}

When we add UV radiation in addition to IR, the resulting winds are stronger and the atmospheric temperature exceeds that of the surface for $\theta>32^\circ$ (solid lines in the bottom panel of Fig.~\ref{fig_PVTold}). The reason for the high atmospheric temperature is the high optical depth in the UV compared to that in the IR.  As shown in the middle panel of Fig.~\ref{fig_epsilon}, UV absorptivity hovers around unity while the IR opacity, being very sensitive to pressure, drops exponentially with $\theta$.

The bottom panel of Fig.~\ref{fig_epsilon} shows that IR absorption from stellar flux and surface blackbody radiation start out as the dominant heating terms, balanced by IR radiative cooling. As pressure drops, however, IR radiation becomes negligible and the UV radiative terms come to dominate the energy balance.

\begin{figure}
	\centering	
	\includegraphics[width=\columnwidth,height=\textheight,keepaspectratio]{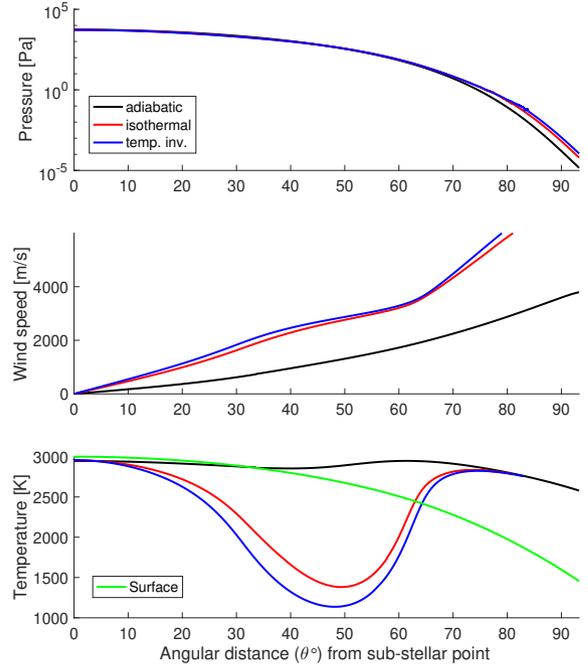}
	\caption{Results of hydrodynamic simulations including IR and UV radiative transfer for different vertical temperature profiles. \emph{Top}: atmospheric pressure. \emph{Middle}: wind velocity. \emph{Bottom}: atmospheric temperature. The black line is the adiabatic profile, red is isothermal, and blue is the inverted temperature profile. Larger $\beta$ increase wind speed and decrease temperature. However, the temperature for all three profiles converge at $\theta>80^\circ$.}
	\label{fig_PVTnew}
\end{figure}

Finally, we incorporate IR and UV radiation to two other vertical temperature profiles: isothermal ($\beta = 1$) and inverted ($\beta = 1.2$). Pressure, wind speed, and temperature for the three vertical profiles are shown in Fig.~\ref{fig_PVTnew}. Higher $\beta$ produce stronger winds and a cooler atmosphere; in Section \ref{subsec_TP}, we offer a physical explanation of this phenomenon. Near the day-night terminator, however, atmospheric temperatures are the same regardless of $\beta$; we discuss this in Section \ref{subsec_UV}.

\section{Discussion}
\label{sec_disc}

\subsection{Impact on dynamics}
\subsubsection{UV radiation}
\label{subsec_UV}

Our simulations show that the exchange of IR radiation between surface and atmosphere does not change the dynamics significantly despite the large fluxes. This is because IR heating and cooling are well balanced, leading to a near zero net radiation budget. However, IR radiation becomes negligible at low pressure making UV stellar flux the dominant heating term, balanced by UV radiative cooling.

\begin{figure}
	\centering	
	\includegraphics[width=\columnwidth,height=\textheight,keepaspectratio]{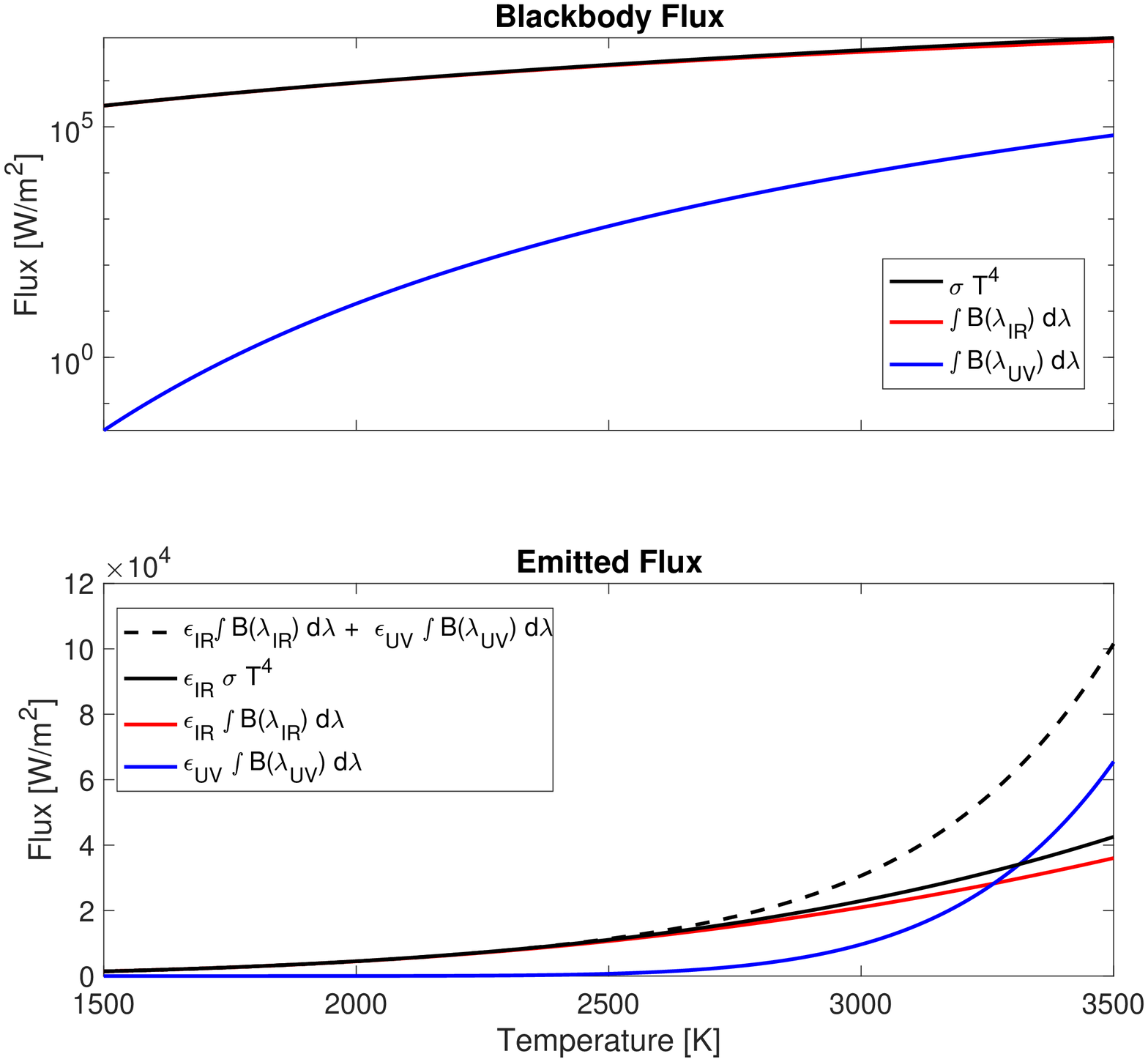}
	\caption{\emph{Top}: atmospheric radiative cooling terms featured in Eq. \ref{5}. The black line is the total blackbody radiative flux across all wavelengths, red is the IR contribution, and blue is the UV contribution. At constant emissivity, IR flux would be order of magnitude greater than UV flux. \emph{Bottom}: emitted blackbody radiation at a fixed pressure. The dashed black line is the total emitted flux, the solid black line is what the emitted flux would be if the total emission is dependent only on IR emissivity, the red line is the individual emitted flux in the IR and blue in the UV. Note that UV emission becomes non-negligible when atmospheric temperatures exceed 2500~K.}
	\label{fig_PlanckInt}
\end{figure}

The importance of UV cooling in a planetary atmosphere is surprising but understandable. Fig. \ref{fig_PlanckInt} shows the IR and UV cooling at a fixed pressure. For uniform emissivity at all wavelengths, the blackbody radiation in the IR accounts for almost all of the radiative cooling. However, because emissivity in the IR is much lower than in the UV, UV emission can overtake the IR with only a slight increase in temperature.

The temperature sensitivity of UV emission  explains the horizontal uniformity in temperature, especially near the terminator ($\theta = 90^\circ$). As pressure drops, UV stellar flux is the only heating term and IR cooling is ineffective. The atmosphere will get hotter until UV radiative cooling takes over. Therefore, temperature will hover around 2600 K where UV heating and cooling are in equilibrium, regardless of the assumed vertical profile (see bottom panel of Fig. \ref{fig_PVTnew}).

\subsubsection{Vertical temperature profile}
\label{subsec_TP}

\begin{figure}
	\centering	
	\includegraphics[width=\columnwidth,height=\textheight,keepaspectratio]{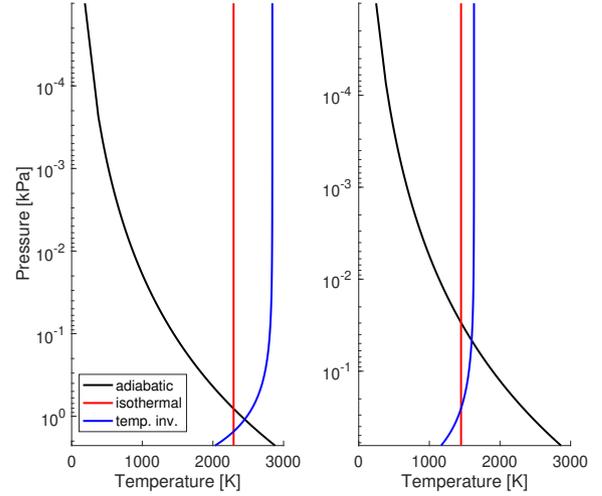}
	\caption{Temperature profiles at $\theta = 30^\circ$ (left) and $\theta = 45^\circ$ (right). While the adiabatic profile stays largely the same at both location (cf. Bottom panel of Fig. \ref{fig_PVTnew}), the isothermal and inverted temperature profiles are cooler at $\theta=45^\circ$.}
	\label{fig_PT2}
\end{figure}

As described in section~\ref{subsec_TBL}, the model atmosphere's vertical temperature profile can be made isothermal by setting $\beta=1$ or inverted for $\beta>1$. This stems from the vertical integration of the pressure which only affects the momentum equation (Eq. \ref{eq_mom}). The higher $\beta$ value implies a much stronger pressure force, the term $\beta C_p T_* P_*/[g r \tan(\theta)]$ in Eq.~\ref{eq_mom}. This is why winds are stronger for isothermal and inverted temperature profiles as seen in Fig. \ref{fig_PVTnew}. We plot the temperature profiles at two different locations in Fig \ref{fig_PT2}.

\begin{figure}
	\centering	
	\includegraphics[width=\columnwidth,height=\textheight,keepaspectratio]{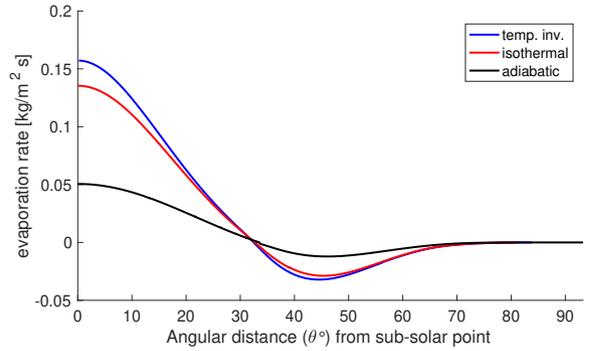}
	\caption{Evaporation rate for different vertical temperature profiles. Because decreasing the lapse rate results in stronger winds, evaporation and condensation are greater for the isothermal and inverted temperature profiles than the adiabatic profile.}
	\label{fig_Eplot}
\end{figure}

Stronger winds also induce more evaporation and condensation and this can be inferred from the mass equation (Eq. \ref{eq_mass}): a larger $V_*$ must lead to a larger $E$. Therefore, a strong vertical temperature inversion has a greater evaporation and condensation rate as seen in Fig. \ref{fig_Eplot}.

Wind speed and atmospheric temperature are coupled within the energy equation (Eq. \ref{eq_en}), hence the increase in wind speed greatly affects temperature. The term $(V^2/2+C_p T_*)$ is balanced by $Q$ which is mostly dependent on the incoming stellar flux. As stellar flux does not change regardless of the vertical temperature profile, a lower temperature is required to balance out the stronger wind speeds.

Radiative transfer models that do not account for horizontal flow predict an inverted temperature profile where the overall atmospheric temperature is hotter than the surface \citep{ito2015theoretical}. The hydrodynamics we present here suggests otherwise: the stronger the inversion, the faster the horizontal flow and the cooler the atmosphere. However, because the lapse rate is set up to be the same everywhere ($\beta$ is constant with respect to $\theta$), the hydrodynamical effects may be overestimated.

\subsection{Impact on observations}
\subsubsection{Transmission spectroscopy}
\label{subsec_transpec}

\begin{figure}
	\centering	
	\includegraphics[width=\columnwidth,height=\textheight,keepaspectratio]{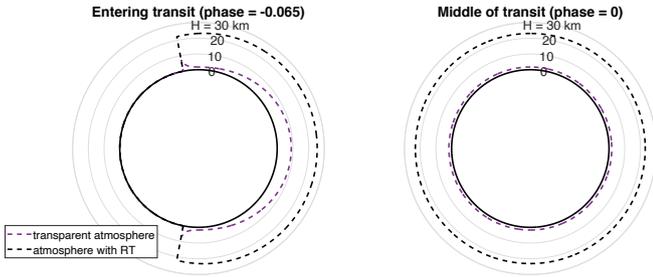}
	\caption{Scale height as observed at ingress (left) and transit (right). Assuming an impact parameter of 0, we plot the expected scale height around the planet's limb. At ingress, the viewing geometry exposes much of the airless nightside (left side of the limb) and we can only see the dayside atmosphere on the right-hand side. At transit, limb measurements only probe where $\theta=90^\circ$ and scale height is constant due to the imposed axial symmetry. Incorporating radiative transfer in our dynamical model produces a much hotter and hence more extended atmosphere near the terminator, leading to a near-constant transmission spectrum throughout the transit.}
	\label{fig_H2}
\end{figure}

We can now use our model atmospheres to simulate observations, the first of which is a transit spectrum. This is done by calculating the transit depth, $D$, the fraction of stellar flux blocked as the planet passes in front of its star. This is evaluated by:
\begin{equation}
    D(\lambda) = \bigg(\frac{r}{r_*}\bigg)^2 + \frac{2}{r_*^2} \int_r^{r_*} b (1-e^{-\tau_d(\lambda,b)}) db,
    \label{eq_transit}
\end{equation}
where $r$ and $r_*$ are the planetary and stellar radii respectively, $b$ is the projected distance to the planet's centre, and $\tau_d$ is the optical depth along the cord. Optical depth is calculated by:
\begin{equation}
    \tau_d(b) = \frac{x P_*}{m g} \sqrt{\frac{2 \pi r} {H}},
\end{equation}
where $H$ is the atmospheric scale height. In the middle of the transit, the relevant surface pressure $P_*$ and $H$ must be evaluated at $\theta=90^\circ$, but the extreme geometry of lava planets means that the a wide range of $\theta$, surface pressures, and scale heights are probed throughout transit, as shown in Fig.~\ref{fig_H2}. Because the boundary layer is geometrically thin compared to scale height, $P_*$ can be approximated as the surface pressure.

Due to the UV radiative equilibrium, atmospheric temperatures converge to the same temperature at the terminator regardless of the vertical temperature profile. 
These temperatures are much hotter than that of the transparent atmosphere described by \cite{nguyen2020modelling}, increasing the atmospheric scale height and providing a stronger signal for transmission spectroscopy of lava planets.

\begin{figure*}
	\centering	
	\includegraphics[width=\textwidth,height=\textheight,keepaspectratio]{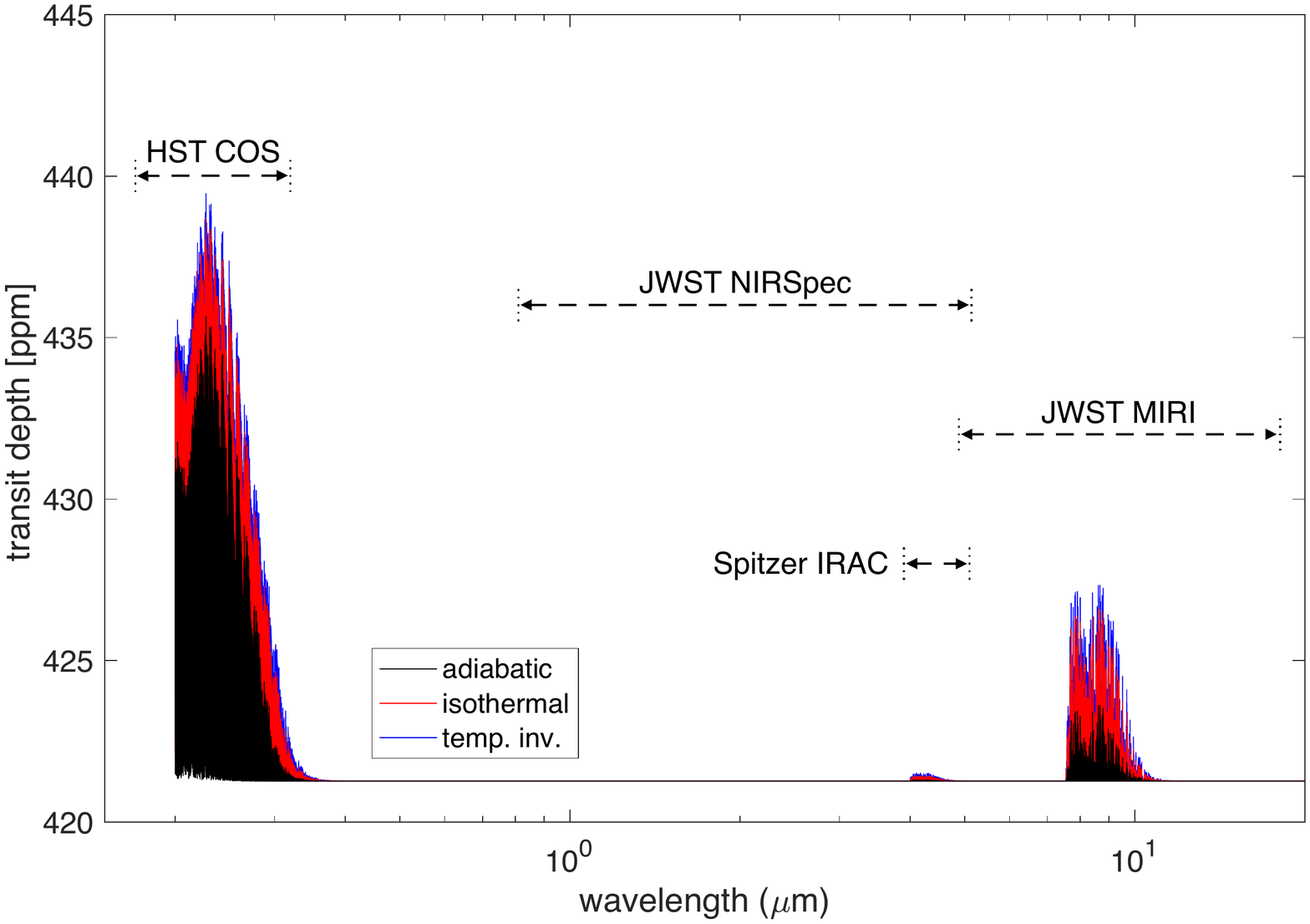}
	\caption{Expected transit spectrum of K2-141b for three possible vertical temperature profiles. Dashed lines show the bandpasses for HST/COS \citep{green2011cosmic}, JWST/NIRSpec \citep{birkmann2016jwst}, Spitzer/IRAC channel 2, and JWST/MIRI \citep{wells2015mid}. The UV spectral feature is in-band for Hubble but is likely too small to be detectable. Likewise, the IR features around 4.5 and 9 $\mu$m, while in-band for Spitzer or JWST, are too faint to be disentangled from noise. Although cases with different T-P profiles have roughly the same scale height at $\theta = 90^\circ$, the surface pressure is slightly larger for profiles with a negative lapse rate. This explains the slight differences in transit depth between the T-P profiles.}
	\label{fig_transspec}
\end{figure*}

Evaluating $\tau_d$ and completing the integration in Eq.~\ref{eq_transit} produces the transit spectra shown in Fig.~\ref{fig_transspec}. 
The biggest peak is at $\sim$0.25 $~\mu$m and the second biggest at $\sim$9$~\mu$m, both corresponding to SiO absorption features since our model atmosphere is entirely composed of that gas. By virtue of having slightly higher atmospheric pressure and therefore greater optical depth at the terminator (see Fig.~\ref{fig_PVTnew}), the isothermal and inverted atmospheres have larger transit spectrum features than the adiabatic case.

The spectral feature in the UV is the strongest but likely requires precision beyond the capabilities of the Hubble Space Telescope. But we only simulate the bound atmosphere of the planet---it is possible that UV observations with Hubble would be sensitive to its exosphere. Transit spectral features in the range of JWST NIRSpec and Spitzer IRAC are negligible while features in the range of JWST MIRI are significant. We use the PandExo tool from \cite{batalha2017pandexo} to simulate JWST measurement uncertainties for 16 transits and conclude that neither NIRSpec or MIRI have the precision to identify any spectral features induced by the atmosphere in transit.




\newpage
\subsubsection{Emission spectroscopy}
\label{subsec_emispec}

\begin{figure}
	\centering	
	\includegraphics[width=\columnwidth,height=\textheight,keepaspectratio]{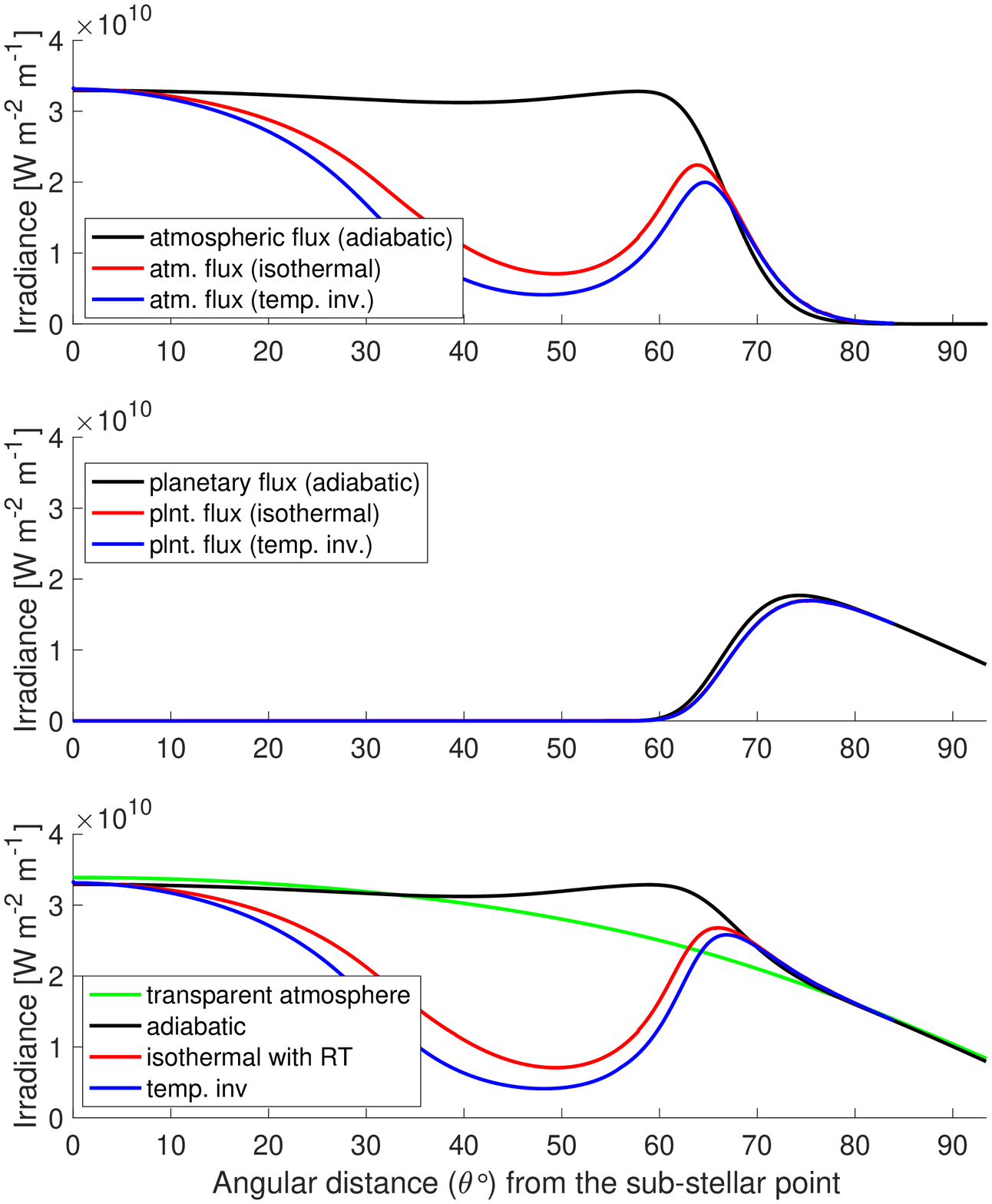}
	\caption{Irradiance of K2-141b at 4.5 $\mu$m i.e, at a wavelength where the atmosphere is optically thick over much of the dayside. \emph{Top}: modeled outgoing atmospheric irradiance. \emph{Middle}: outgoing planetary irradiance. \emph{Bottom}: total outgoing irradiance at the top of the atmosphere. The adiabatic case has stronger emission at this wavelength while the other two cases have a weaker emission than the transparent atmosphere.}
	\label{fig_OutIrr}
\end{figure}

To calculate the emission spectrum of K2-141b at eclipse, we first calculate the top-of-atmosphere outgoing irradiance, which is a combination of atmospheric and planetary emission. While the greenhouse effect barely changes the surface temperature, it is significant enough to affect the planet's emission spectrum. Emission from the atmosphere is calculated as $\epsilon_\lambda B(\lambda,T_*)$ while the surface contribution is calculated as $(1-\epsilon_\lambda) B(\lambda,T_s)$. The top two panels of Fig.~\ref{fig_OutIrr} show the atmospheric and surface irradiance at 4.5~$\mu$m, where there is a strong SiO spectral feature. The bottom panel shows the total top-of-atmosphere outgoing irradiance.

\begin{figure}
	\centering	
	\includegraphics[width=\columnwidth,height=\textheight,keepaspectratio]{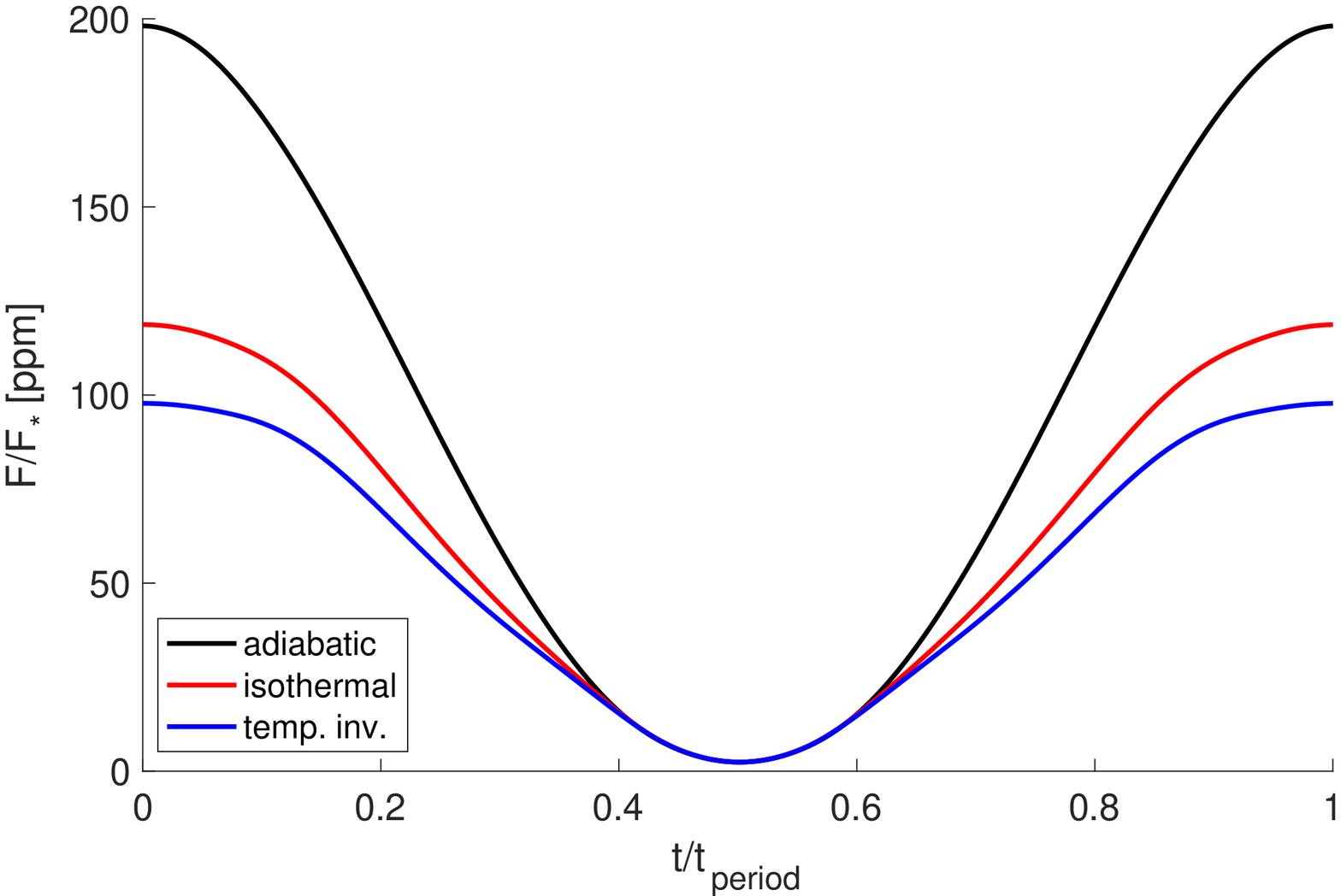}
	\caption{Simulated 4.5 $\mu$m phase curves from the three different vertical temperature structures. Note that eclipses have been omitted at phases 0 and 1; transit is omitted at phase 0.5. The adiabatic temperature profile produced an atmosphere with stronger dayside emission and therefore a larger phase amplitude.}
	\label{fig_Phase}
\end{figure}

\begin{figure*}
	\centering	
	\includegraphics[width=\textwidth,height=\textheight,keepaspectratio]{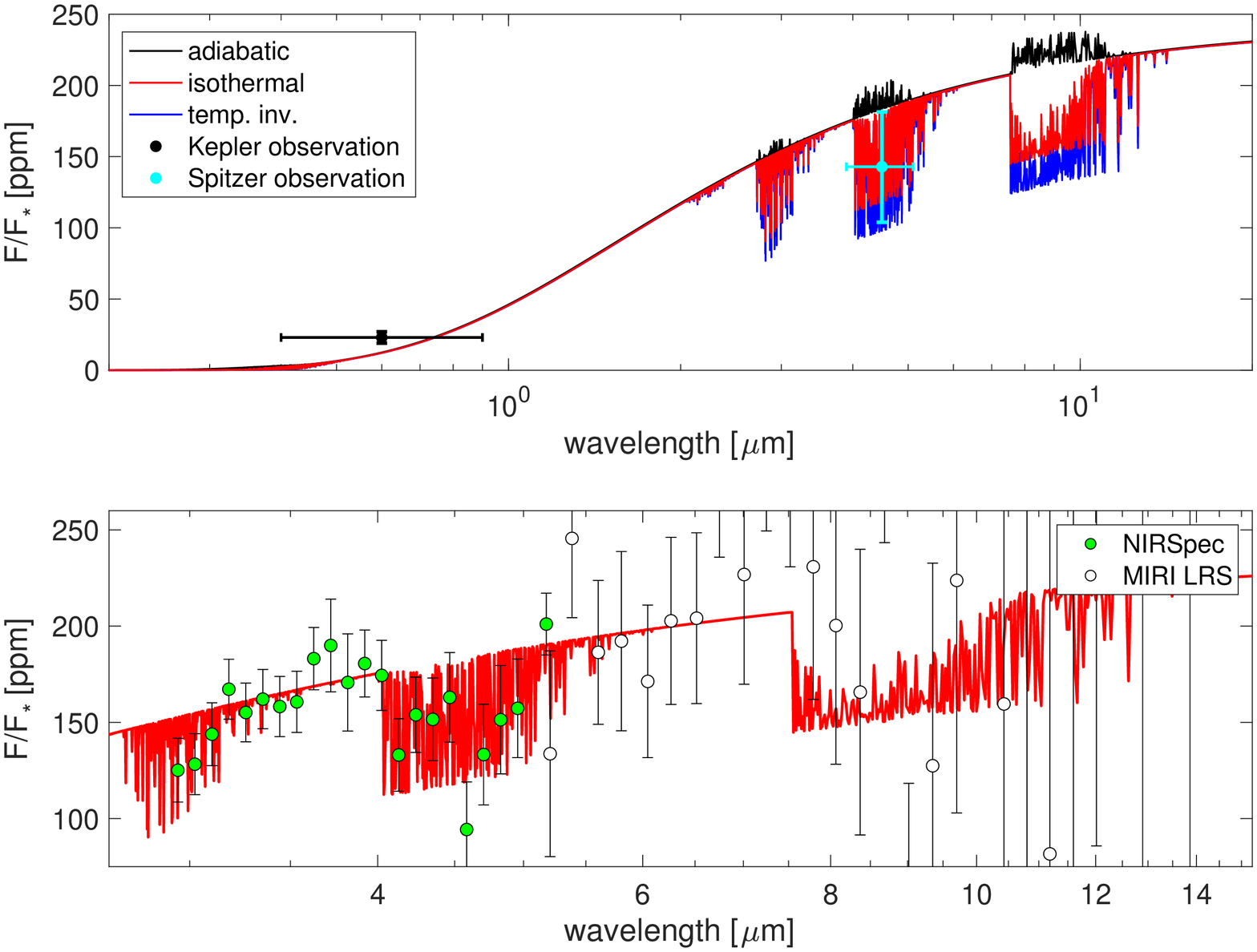}
	\caption{\emph{Top}: simulated eclipse spectrum of K2-141b for three different assumed vertical temperature pressure profiles. The black dot and its bars show K2-141b's occultation depth as measured by Kepler \citep{malavolta2018ultra}, as well as Kepler's bandwidth \citep{van2016kepler}. The cyan dot and its bars likewise show the depth as measured by Spitzer \citep{zieba2022k2}. The adiabatic case exhibits SiO emission features while the vertically isothermal and inverted models have absorption features. This counter-intuitive behaviour originates from the relation between vertical temperature profile, wind speed, and overall atmospheric temperature with respect to the surface. \emph{Bottom}: simulated JWST observations at eclipse. We take the spectrum from the isothermal profile case and simulate NIRSpec and MIRI measurements for 16 eclipses using the tool PandExo \citep{batalha2017pandexo}. Results for NIRSpec are promising and can detect the SiO absorption feature at 4--5 $\mu$m.}
	\label{fig_eclispec}
\end{figure*}

The emission spectrum is a function of orbital phase because different regions of the planet rotate in and out of view. Therefore, outgoing irradiance with respect to $\theta$ is translated to a spherical coordinates with co-latitude ($\alpha$) and longitude ($\phi$). We then integrate the irradiance over the dayside hemisphere following \cite{2008ApJ...678L.129C} to find the emission, $I_e$:
\begin{equation}
    I_e(\lambda,\zeta) = \int_0^\pi \int_{\zeta-\pi/2}^{\zeta+\pi/2} I(\alpha,\phi,\lambda) \sin^2(\alpha) \cos(\phi-\zeta) d\phi d\alpha,
\end{equation}


\noindent where $\zeta$ is the position of the planet in its orbit: $\zeta=0$ at superior conjunction and $\zeta=\pi$ at inferior conjunction. Keeping $\lambda$ fixed at 4.5~$\mu$m and calculating $I_e(\zeta)$ yields the phase curve shown in Fig. \ref{fig_Phase}. Because the dayside emission from the isothermal and inverted temperature profiles are smaller than the adiabatic case, their phase amplitudes are likewise smaller. Eclipse occurs when $\zeta = 0$, and calculating $I_e$ with respect to $\lambda$ at this point leads to the eclipse spectrum plotted in Fig.~\ref{fig_eclispec}. 

Paradoxically, the adiabatic profile results in SiO emission features while the isothermal and inverted profiles result in absorption features. SiO spectral features probe the atmosphere rather than the surface over much of the dayside.  
The adiabatic profile produces an atmosphere hotter than the surface and therefore its spectrum shows emission features. Meanwhile, the isothermal and inverted temperature profiles lead to cooler atmospheres (bottom panel of Fig. \ref{fig_PVTnew}) and hence their spectra show absorption features. 

We simulate JWST measurement uncertainties for 16 eclipses using PandExo \citep{batalha2017pandexo}, plotted in the bottom panel of Fig. \ref{fig_eclispec}. Spectral features at 4.5 $\mu$m can be detected with NIRSpec. Spectral features at longer wavelengths are difficult to see with MIRI but observing more eclipses or phase curves may reduce the errors enough to make out the features.

\newpage
\section{Conclusions}
\label{sec_conc}

We have improved upon the 1D hydrodynamic model of \cite{nguyen2020modelling} by implementing radiative transfer. We applied this new model to the lava planet K2-141b and tested the effects of different vertical temperature profiles. Our analysis leads to two key insights.

Our first insight is that the atmosphere far from the sub-stellar point will be hotter than the surface. Since UV heating dominates over IR when pressure is low, atmospheric temperatures will rise to $\sim$2600K, at which point UV radiative cooling takes effect. As a result, atmospheric temperatures should be horizontally uniform near the terminator ($\theta \approx 90^\circ$), regardless of vertical temperature profile.

The second insight is that a vertical temperature inversion produces a cooler atmosphere. A strong temperature inversion also leads to a greater horizontal pressure gradient force and hence greater wind speeds. To conserve energy, thermal energy in the atmosphere is reduced as a reaction to the greater kinetic energy. This result is particularly important as it shows how hydrodynamics can lead to a significantly different atmosphere when compared to models that focus on 1D radiative balance \citep[e.g.,][]{ito2015theoretical}.

We also simulated the transit spectrum of K2-141b (Fig.~\ref{fig_transspec}) and found that UV radiation leads to a much bigger scale height at the terminator than what was previously predicted \citep{nguyen2020modelling}. However, the surface pressure is very small and so features in the IR are too subtle for Spitzer or JWST to detect, while the larger UV SiO feature likely have to wait for a more sensitive spectrograph than is currently available on Hubble.

Our simulated eclipse spectra of K2-141b (Fig.~\ref{fig_eclispec}) are more promising, however. Because the different vertical temperature profiles lead to different dayside atmospheric temperatures, emission spectra ought to be distinguishable with Spitzer and JWST. The adiabatic profile has a warmer atmosphere so its spectrum has significant SiO emission features. The isothermal and inverted vertical temperature profiles, on the other hand, produce atmospheres cooler than the surface and hence SiO absorption features. Emission spectra are not only useful for detecting the atmosphere, but can also help to infer more nuanced properties such as vertical temperature profile.


\section*{Acknowledgements}
This work was made possible by the Natural Science and Engineering Research Council (NSERC) of Canada’s Collaborative Research  and  Training  Experience  Program  (CREATE)  for  Technology for Exo-Planetary Sciences (TEPS). This work was also made possible by the NASA XRP grant NNX17AC02G. This work has made use of the MUSCLES Treasury Survey High-Level Science Products. NBC acknowledges support  from  the  McGill  Space  Institute (MSI) and l’Institut de recherche sur les exoplan\`etes (iREx). 

\section*{Data Availability}

Data is available upon request.


\newpage
\bibliographystyle{mnras}
\bibliography{manuscript}

\begin{thebibliography}{}
\makeatletter
\relax
\def\mn@urlcharsother{\let\do\@makeother \do\$\do\&\do\#\do\^\do\_\do\%\do\~}
\def\mn@doi{\begingroup\mn@urlcharsother \@ifnextchar [ {\mn@doi@}
  {\mn@doi@[]}}
\def\mn@doi@[#1]#2{\def\@tempa{#1}\ifx\@tempa\@empty \href
  {http://dx.doi.org/#2} {doi:#2}\else \href {http://dx.doi.org/#2} {#1}\fi
  \endgroup}
\def\mn@eprint#1#2{\mn@eprint@#1:#2::\@nil}
\def\mn@eprint@arXiv#1{\href {http://arxiv.org/abs/#1} {{\tt arXiv:#1}}}
\def\mn@eprint@dblp#1{\href {http://dblp.uni-trier.de/rec/bibtex/#1.xml}
  {dblp:#1}}
\def\mn@eprint@#1:#2:#3:#4\@nil{\def\@tempa {#1}\def\@tempb {#2}\def\@tempc
  {#3}\ifx \@tempc \@empty \let \@tempc \@tempb \let \@tempb \@tempa \fi \ifx
  \@tempb \@empty \def\@tempb {arXiv}\fi \@ifundefined
  {mn@eprint@\@tempb}{\@tempb:\@tempc}{\expandafter \expandafter \csname
  mn@eprint@\@tempb\endcsname \expandafter{\@tempc}}}

\bibitem[\protect\citeauthoryear{Barton, Yurchenko  \& Tennyson}{Barton
  et~al.}{2013}]{barton2013exomol}
Barton E.~J.,  Yurchenko S.~N.,   Tennyson J.,  2013, Monthly Notices of the
  Royal Astronomical Society, 434, 1469

\bibitem[\protect\citeauthoryear{Batalha et~al.,}{Batalha
  et~al.}{2011}]{batalha2011kepler}
Batalha N.~M.,  et~al., 2011, The Astrophysical Journal, 729, 27

\bibitem[\protect\citeauthoryear{Batalha et~al.,}{Batalha
  et~al.}{2017}]{batalha2017pandexo}
Batalha N.~E.,  et~al., 2017, Publications of the Astronomical Society of the
  Pacific, 129, 064501

\bibitem[\protect\citeauthoryear{Birkmann et~al.,}{Birkmann
  et~al.}{2016}]{birkmann2016jwst}
Birkmann S.~M.,  et~al., 2016, in Space Telescopes and Instrumentation 2016:
  Optical, Infrared, and Millimeter Wave. p. 99040B

\bibitem[\protect\citeauthoryear{Castan \& Menou}{Castan \&
  Menou}{2011}]{castan2011atmospheres}
Castan T.,  Menou K.,  2011, The Astrophysical Journal Letters, 743, L36

\bibitem[\protect\citeauthoryear{{Cowan} \& {Agol}}{{Cowan} \&
  {Agol}}{2008}]{2008ApJ...678L.129C}
{Cowan} N.~B.,  {Agol} E.,  2008, \mn@doi [\apjl] {10.1086/588553}, \href
  {https://ui.adsabs.harvard.edu/abs/2008ApJ...678L.129C} {678, L129}

\bibitem[\protect\citeauthoryear{Dang et~al.,}{Dang
  et~al.}{2021}]{dang2021hell}
Dang L.,  et~al., 2021, JWST Proposal. Cycle 1, p.~2347

\bibitem[\protect\citeauthoryear{Essack, Seager  \& Pajusalu}{Essack
  et~al.}{2020}]{essack2020low}
Essack Z.,  Seager S.,   Pajusalu M.,  2020, The Astrophysical Journal, 898,
  160

\bibitem[\protect\citeauthoryear{France et~al.,}{France
  et~al.}{2016}]{france2016muscles}
France K.,  et~al., 2016, The Astrophysical Journal, 820, 89

\bibitem[\protect\citeauthoryear{Gladstone \& Young}{Gladstone \&
  Young}{2019}]{gladstone2019new}
Gladstone G.~R.,  Young L.~A.,  2019, Annual Review of Earth and Planetary
  Sciences, 47, 119

\bibitem[\protect\citeauthoryear{Gordon et~al.,}{Gordon
  et~al.}{2017}]{gordon2017hitran2016}
Gordon I.~E.,  et~al., 2017, Journal of Quantitative Spectroscopy and Radiative
  Transfer, 203, 3

\bibitem[\protect\citeauthoryear{Green et~al.,}{Green
  et~al.}{2011}]{green2011cosmic}
Green J.~C.,  et~al., 2011, The Astrophysical Journal, 744, 60

\bibitem[\protect\citeauthoryear{Ingersoll, Summers  \& Schlipf}{Ingersoll
  et~al.}{1985}]{ingersoll1985supersonic}
Ingersoll A.~P.,  Summers M.~E.,   Schlipf S.~G.,  1985, Icarus, 64, 375

\bibitem[\protect\citeauthoryear{Ito \& Ikoma}{Ito \&
  Ikoma}{2021}]{ito2021hydrodynamic}
Ito Y.,  Ikoma M.,  2021, Monthly Notices of the Royal Astronomical Society,
  502, 750

\bibitem[\protect\citeauthoryear{Ito, Ikoma, Kawahara, Nagahara, Kawashima  \&
  Nakamoto}{Ito et~al.}{2015}]{ito2015theoretical}
Ito Y.,  Ikoma M.,  Kawahara H.,  Nagahara H.,  Kawashima Y.,   Nakamoto T.,
  2015, The Astrophysical Journal, 801, 144

\bibitem[\protect\citeauthoryear{Jolicard, Zucconi, Drira, Spielfieldel  \&
  Feautrier}{Jolicard et~al.}{1997}]{jolicard1997photodissociation}
Jolicard G.,  Zucconi J.-M.,  Drira I.,  Spielfieldel A.,   Feautrier N.,
  1997, The Journal of chemical physics, 106, 10105

\bibitem[\protect\citeauthoryear{Kite, Fegley~Jr, Schaefer  \& Gaidos}{Kite
  et~al.}{2016}]{kite2016atmosphere}
Kite E.~S.,  Fegley~Jr B.,  Schaefer L.,   Gaidos E.,  2016, The Astrophysical
  Journal, 828, 80

\bibitem[\protect\citeauthoryear{Kopal}{Kopal}{1954}]{kopal1954photometric}
Kopal Z.,  1954, Monthly Notices of the Royal Astronomical Society, 114, 101

\bibitem[\protect\citeauthoryear{Kurucz}{Kurucz}{1992}]{kurucz1992atomic}
Kurucz R.~L.,  1992, Revista Mexicana de Astronomia y Astrofisica, vol. 23, 23

\bibitem[\protect\citeauthoryear{L{\'e}ger et~al.,}{L{\'e}ger
  et~al.}{2011}]{leger2011extreme}
L{\'e}ger A.,  et~al., 2011, Icarus, 213, 1

\bibitem[\protect\citeauthoryear{Malavolta et~al.,}{Malavolta
  et~al.}{2018}]{malavolta2018ultra}
Malavolta L.,  et~al., 2018, The Astronomical Journal, 155, 107

\bibitem[\protect\citeauthoryear{Miguel, Kaltenegger, Fegley  \&
  Schaefer}{Miguel et~al.}{2011}]{miguel2011compositions}
Miguel Y.,  Kaltenegger L.,  Fegley B.,   Schaefer L.,  2011, The Astrophysical
  Journal Letters, 742, L19

\bibitem[\protect\citeauthoryear{Nguyen, Cowan, Banerjee  \& Moores}{Nguyen
  et~al.}{2020}]{nguyen2020modelling}
Nguyen T.~G.,  Cowan N.~B.,  Banerjee A.,   Moores J.~E.,  2020, Monthly
  Notices of the Royal Astronomical Society, 499, 4605

\bibitem[\protect\citeauthoryear{Schaefer \& Fegley}{Schaefer \&
  Fegley}{2009}]{schaefer2009chemistry}
Schaefer L.,  Fegley B.,  2009, The Astrophysical Journal Letters, 703, L113

\bibitem[\protect\citeauthoryear{Tsang, Spencer, Lellouch, Lopez-Valverde  \&
  Richter}{Tsang et~al.}{2016}]{tsang2016collapse}
Tsang C.~C.,  Spencer J.~R.,  Lellouch E.,  Lopez-Valverde M.~A.,   Richter
  M.~J.,  2016, Journal of Geophysical Research: Planets, 121, 1400

\bibitem[\protect\citeauthoryear{Van~Cleve \& Caldwell}{Van~Cleve \&
  Caldwell}{2016}]{van2016kepler}
Van~Cleve J.~E.,  Caldwell D.~A.,  2016, Kepler Science Document
  KSCI-19033-002, p.~1

\bibitem[\protect\citeauthoryear{Wells et~al.,}{Wells
  et~al.}{2015}]{wells2015mid}
Wells M.,  et~al., 2015, Publications of the Astronomical Society of the
  Pacific, 127, 646

\bibitem[\protect\citeauthoryear{Widger~Jr \& Woodall}{Widger~Jr \&
  Woodall}{1976}]{widger1976integration}
Widger~Jr W.,  Woodall M.,  1976, Bulletin of the American Meteorological
  Society, 57, 1217

\bibitem[\protect\citeauthoryear{Wordsworth}{Wordsworth}{2015}]{wordsworth2015atmospheric}
Wordsworth R.,  2015, The Astrophysical Journal, 806, 180

\bibitem[\protect\citeauthoryear{Zieba et~al.,}{Zieba
  et~al.}{2022}]{zieba2022k2}
Zieba S.,  et~al., 2022, arXiv preprint arXiv:2203.00370

\makeatother
\end{thebibliography}



\appendix

\section{Definite integral of Planck function}

Evaluating the radiative cooling requires a definite integral of the Planck function (Eq.~\ref{5}). Doing the integral through finite differences can be computationally expensive but \cite{widger1976integration} provide a better alternative. We start with the definite integral of the Planck function from a wavelength $\lambda_1$ to infinity:

\begin{equation}
    \int_{\lambda_1}^{\infty} \frac{2 h c}{\lambda^5} \frac{1}{e^(hc/\lambda k T) - 1} = -\frac{2 \pi k^4 T^4}{h^3 c^2} \sum_n G e^{-n J},
\end{equation}

\noindent where $G = J^3/n + 3J^2/n^2 + 6J/n^3 + 6/n^4$ and $J = hc/(\lambda k T)$. Therefore, the definite integral between two wavelengths is:

\begin{equation}
    \int_{\lambda_1}^{\lambda_2} (...) = \int_{\lambda_1}^{\infty} (...) - \int_{\lambda_2}^{\infty} (...),
\end{equation}

\noindent and each term is calculated via the summation provided.

\bsp	
\label{lastpage}
\end{document}